\documentclass{article}
\usepackage{arxiv}
\usepackage{color}
\usepackage{subfigure}
\usepackage{graphicx}
\usepackage{dcolumn}
\usepackage{bm}
\usepackage{floatrow}
\usepackage{amsmath}
\usepackage{makeidx}
\usepackage{epsfig}
\usepackage{verbatim}
\usepackage{color}
\usepackage{subfigure}
\usepackage{graphicx}
\usepackage{dcolumn}
\usepackage{bm}
\usepackage{floatrow}
\usepackage{amsmath}
\usepackage{makeidx}
\usepackage{epsfig}
\usepackage{amsfonts}
\usepackage{hyperref}
\usepackage{array}
\usepackage{epstopdf}
\usepackage{braket}
\usepackage{multirow}
\newcommand{\RNum}[1]
{\uppercase\expandafter{\romannumeral #1\relax}}
\newcommand{\Rnum}[1]
{\lowercase\expandafter{\romannumeral #1\relax}}
\usepackage{array}
\usepackage{epstopdf}
\usepackage{multirow}
\usepackage[utf8]{inputenc} 
\usepackage[T1]{fontenc}    
\usepackage{hyperref}       
\usepackage{url}            
\usepackage{booktabs}       
\usepackage{amsfonts}       
\usepackage{nicefrac}       
\usepackage{microtype}      
\usepackage{lipsum}		
\title{Temperature as perturbation in quantum mechanics}
\author{
  Ashkan Shekaari \\
  Department of Physics\\
  K. N. Toosi University of Technology\\
  Tehran, 15875-4416, Iran \\
  \texttt{shekaari@email.kntu.ac.ir} \\
   \And
 Mahmoud Jafari \thanks{Corresponding author} \\
  Department of Physics\\
  K. N. Toosi University of Technology\\
  Tehran, 15875-4416, Iran \\
  \texttt{jafari@kntu.ac.ir} \\}
\begin{document}
\maketitle
\begin{abstract}
The perturbative approach was adopted to develop a temperature-dependent version of non-relativistic quantum mechanics in the limit of low-enough temperatures. A generalized, self-consistent Hamiltonian was therefore constructed for an arbitrary quantum-mechanical system in a way that the ground-state Hamiltonian turned out to be just a limiting case at absolute zero. The weak-coupling term connecting the system of interest and its immediate environment was accordingly treated as the perturbation. Applying the obtained generalized Hamiltonian to some typical quantum systems with exact zero-temperature solutions, including the free particle in a box, the free particle in vacuum, and the harmonic oscillator, up to the first order of self-consistency, therefore corrected their associated Hamiltonians, energy spectrums, and wavefunctions to be consistent with the low-temperature limit. Further investigation revealed some kind of quantum tunneling effect by a residual probability for the free particle in a box, as a chief consequence of thermally coupling to the reservoir. The possible effects of thermal environment on the main properties of the wavefunctions were also thoroughly examined and discussed.
\end{abstract}
\section{Introduction}
Temperature is a collective phenomenon and is therefore only tangible in dealing with macroscopic systems containing as large number of microscopic particles as several orders of magnitude of the Avogadro's number. Nonetheless, statistical mechanics~\cite{1} successfully defines absolute temperature of a given system at the microscopic level in terms of its number of accessible microstates, similar to its forerunner, the kinetic theory of gases~\cite{2}, providing a microscopic expression for instantaneous temperature of the system based on the movements of the constituent particles. The question that naturally arises is therefore whether it is possible to assign a temperature to a quantum-mechanical system, or more technically, how do the quantum-mechanical properties of the system of interest evolve at finite temperatures compared to those at absolute zero?

Historically, studying finite-temperature behaviors of physical systems has indeed been a subject of thermodynamics. For quantum-mechanical systems, nonetheless, one has to deploy statistical mechanics based on the fact that thermodynamics is an inherently macroscopic theory. However, the formulation of quantum statistics as the rewriting of statistical mechanics in terms of operators and wavefunctions has not yet introduced any new physical idea as such; rather, it has provided us with a highly-suited tool for inquiring into typical quantum systems, without taking the explicit approach of constructing a generalized, temperature-dependent Hamiltonian.

So far, many efforts have been devoted to dealing with temperature at scales far below the thermodynamic limit. Accurate temperature reading at the nanoscale has found its applications in several areas ranging from quantum thermodynamics~\cite{3,4,5,6,7,8}, to materials science~\cite{9,10,11,12}, to medicine and biology~\cite{13,14}, in which controlling the performance of quantum thermal systems and devices is of foremost importance~\cite{15}. Interest in this field, in fact, has been motivated by the recent advancements in nanoscale thermometry~\cite{16,17,18,19}, such as carbon nanothermometers~\cite{20,21}, diamond sensors~\cite{22}, and scanning thermal microscopes~\cite{23,24,25,26,27,28}. The current technological developments have provided us with realizing extremely small thermometers~\cite{29,30,31} to overcome the challenging issue of controlling thermal behaviors of physical systems at the micro and nano scales~\cite{32}. De Pasquale {\em{et al.}}~\cite{33} have introduced a hypothetical quantifier called "local quantum thermal susceptibility" for the best achievable accuracy in temperature estimation via local measurements, which relies on the basic concepts of quantum estimation theory~\cite{34,35}. Such a quantifier, in fact, provides an operative strategy to address the local thermal response of a given quantum system at equilibrium, also highlights the local distinguishability of the ground state from the excited sub-manifolds. Miller and Anders~\cite{36}---based on the fact that for nanoscale systems, deviations from classical thermodynamics arise due to their interactions with the environment---have accordingly derived a generalized thermodynamic uncertainty relation, which is valid for both classical and quantum-mechanical systems at all coupling strengths by taking into account such interactions within the framework of quantum estimation theory. They also showed that the non-commutativity between the state and the effective energy operator of the system results in quantum fluctuations that increase temperature uncertainty~\cite{37}. A set of methods to calculate expectation values of physical observables of quantum-mechanical systems at finite temperatures has also been developed under the title of "thermal quantum field theory"~\cite{38,39,40}, which exploit the concept of imaginary time to establish connection between quantum and statistical mechanics.

In the present work, we adopted a different, perturbative approach~\cite{41} to develop a temperature-dependent version of non-relativistic quantum mechanics, in order to investigate the behavior of quantum-mechanical systems at finite and low-enough temperatures. A generalized, temperature-dependent Hamiltonian was accordingly derived, which contains a weak coupling term connecting the system of interest and its thermal environment. The equilibrium temperature of the system was also treated as a mere parameter, rather than as a quantum-mechanical observable. We finally applied the generalized Hamiltonian to some typical quantum systems with exact ground-state solutions, including the free particle in a box, the free particle in vacuum, and the harmonic oscillator, and the obtained results and consequences were also thoroughly discussed. Sec.~\ref{sec:2} further illustrates the underlying assumptions of our formalism.
\section{Assumptions}
\label{sec:2}
\begin{figure}
	\centering
	\fbox{\rule[0cm]{0cm}{0cm} \rule[0cm]{0cm}{0cm} 
	\includegraphics[scale=0.12]{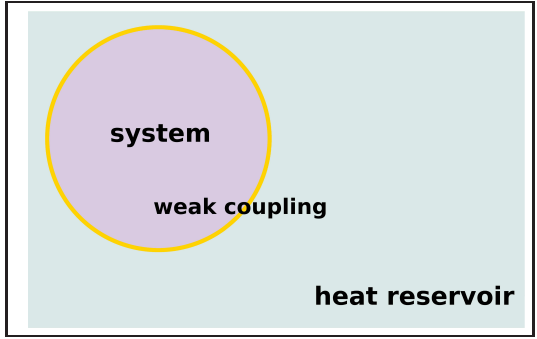}}
	\caption{
		Depiction of a closed system, which is weakly coupled to a much larger heat reservoir, forming together a single isolated super-system as well---rendered in Gimp (version 2.8)~\cite{45}.}
	    \label{fig:1}
\end{figure}

The system of interest, which is assumed to be closed, was prepared in a state of thermal equilibrium by placing it in a weak thermal contact with a much larger heat reservoir at temperature $T$ (Fig.~\ref{fig:1}), in a way that the "system $+$ reservoir", in turn, is an isolated super-system. In summary, (i) the equilibrium temperature of the system is treated as a parameter, rather than as an observable; (ii) the system--reservoir coupling takes place so that the mutual equilibrium state can then be best described by the canonical ensemble; (iii) the strength of the system--reservoir interaction is assumed to be so negligible as the local equilibrium state of the system will accordingly be of the Gibbs form (this is the weak-coupling assumption)~\cite{42,43}; (iv) the internal energy $(U)$ of the system can be determined by the bare Hamiltonian of the weakly-coupled system of interest~\cite{44}; (v) the equilibrium temperature of the system is so low that the perturbative approach is indeed applicable and valid (the effect of finite temperatures therefore emerge as corrections to the corresponding zero-temperature Hamiltonian); and (vi) the state of the system changes during reversible processes so that the inexact differentials of heat ($\delta Q$) and work ($\delta W$) in the first law of thermodynamics can be replaced by the exact ones.
\section{\label{sec:3}Analytic method}
The internal energy $(U)$ of the system containing $N$ particles is given by
\begin{equation}
\label{eq:eq1}
U=\sum_{i=1}^{N}\Big[T_{i}+v(\mathrm{{\bf{r}}_{i}})\Big]+\sum_{i=1}^{N}\sum_{j>i}^{N}\frac{1}{|\mathrm{{\bf{r}}_{i}}-\mathrm{{\bf{r}}_{j}}|},
\end{equation}
where $T_{i}$ is the kinetic energy of particle $i$, $v(\mathrm{{\bf{r}}_{i}})$ is the potential at $\mathrm{{\bf{r}}_{i}}$, and the last term is the interaction between particles $i$ and $j$ located at $\mathrm{{\bf{r}}_{i}}$ and $\mathrm{{\bf{r}}_{j}}$, respectively (bold characters denote vectors). From the first law of thermodynamics, we have
\begin{equation}
\label{eq:eq2}
dU+\delta Q=\delta W\Longrightarrow U+Q=W,
\end{equation}
where $dU$ is change in the internal energy of the system, $\delta Q$ is the net amount of heat supplied to the system, and $\delta W$ is consequently the work done by the system. From the second law of thermodynamics, for a closed system that allows the entry or exit of energy at the equilibrium temperature $T$, change in the entropy $(S)$ due to an infinitesimal transfer of heat ($\delta Q$) is then given by
\begin{equation}
\label{eq:eq3}
\delta Q=TdS=d(TS)\Longrightarrow Q=TS.
\end{equation}
Inserting Eqs.~\ref{eq:eq1} and~\ref{eq:eq3} into Eq.~\ref{eq:eq2}, we obtain that
\begin{equation}
\label{eq:eq4}
\sum_{i=1}^{N}\Big[T_{i}+v(\mathrm{{\bf{r}}_{i}})\Big]+\sum_{i=1}^{N}\sum_{j>i}^{N}\frac{1}{|\mathrm{{\bf{r}}_{i}}-\mathrm{{\bf{r}}_{j}}|}+TS=W.
\end{equation}
Using $A(T)=-k_{B}T\ln{\mathcal{Q}}(T)$ as the Helmholtz free energy of the system, it follows that
\begin{equation}
\label{eq:eq5}
S=-\left(\frac{\partial A}{\partial T}\right)_{N,V}=k_{B}\Big(1+T\partial_{T}\Big)\ln {\mathcal{Q}}(T),
\end{equation}
where ${\mathcal{Q}}(T)$ is the associated canonical partition function, and $k_{B}$ is Boltzmann's constant. Plugging Eq.~\ref{eq:eq5} back into Eq.~\ref{eq:eq4} then gives
\begin{eqnarray}
\label{eq:eq6}
\sum_{i=1}^{N}\Big[T_{i}+v(\mathrm{{\bf{r}}_{i}})\Big]+\sum_{i=1}^{N}\sum_{j>i}^{N}\frac{1}{|\mathrm{{\bf{r}}_{i}}-\mathrm{{\bf{r}}_{j}}|}+k_{B}T\Big(1+T\partial_{T}\Big)\ln{\mathcal{Q}}(T)=W.
\end{eqnarray}
Using the corresponding quantum-mechanical operator $(-\hbar^{2}/2m)\nabla_{i}^{2}$ for the kinetic energy $T_i$, and then multiplying both sides of Eq.~\ref{eq:eq6} by the temperature-dependent wavefunction $\mathrm{\Psi}(T,{\bf{r}})$, it is therefore obtained that
\begin{eqnarray}
\label{eq:eq7}
\left[\sum_{i=1}^{N}\left(-\frac{\hbar^{2}}{2m_{i}}\nabla_{i}^{2}+v(\mathrm{{\bf{r}}_{i}})\right)+\sum_{i=1}^{N}\sum_{j>i}^{N}\frac{1}{|\mathrm{{\bf{r}}_{i}}-\mathrm{{\bf{r}}_{j}}|}+k_{B}T\Big(1+T\partial_{T}\Big)\ln {\mathcal{Q}}(T)\right]\mathrm{\Psi}(T,{\bf{r}})=W\mathrm{\Psi}(T,{\bf{r}}),
\end{eqnarray}
where $\hbar$ is Planck's constant divided by $2\pi$, $m_{i}$ is the mass of particle $i$, and the first three terms on the left hand, together, form the zero-temperature Hamiltonian ($\hat{H}_0$) of the system (hat denotes operator). Eq.~\ref{eq:eq7} is indeed the generalized, time-independent Schr{\"o}dinger equation of the many-body system at the equilibrium temperature $T$. Using ${\mathcal{Q}}(T)={\mathrm{Tr}}(e^{-\beta\hat{H}})$, the total Hamiltonian in Eq.~\ref{eq:eq7} accordingly takes the form
\begin{equation}
\label{eq:eq8}
\hat{H}=\hat{H}_{0}+k_{B}T\Big(1+T\partial_{T}\Big)\ln{\mathrm{Tr}}\left(e^{-\beta\hat{H}}\right),
\end{equation}
which, in fact, is a self-consistent Hamiltonian because of $\hat{H}$ being appeared on both sides. The more illustrative expression of Eq.~\ref{eq:eq8} reflecting its self-consistency is
\begin{equation}
\label{eq:eq9}
\hat{H_{I}}=\hat{H}_{0}+k_{B}T\ln{\mathrm{Tr}}\left(e^{-\beta\hat{H}_{I-1}}\right),
\end{equation}
where $I$ $(\ge1)$ denotes the order of self-consistency, and the quadratic term in $T$ in Eq.~\ref{eq:eq8} has also been dropped based on the previous assumption of being in the limit of low-enough temperatures. Eq.~\ref{eq:eq9} lies at the very heart of the present approach. Starting from $I=1$, which produces $\hat{H}_{0}$ on the right side of Eq.~\ref{eq:eq9}, any order $I$ $(>0)$ of $H_{I}$ could then be calculated. Here, we proceed by considering only the case $I=1$ based on the fact that it results in the largest $E_{p}(T)$, which is not as such ignorable. As a result, the spectrum of the generalized Hamiltonian would be
\begin{equation}
\label{eq:eq92}
E(T)=E_{gs}(0)+k_{B}T\ln{\mathrm{Tr}}\Big(e^{-\beta E(0)}\Big),
\end{equation}
where $k_{B}T\ln{\mathrm{Tr}}\big(e^{-\beta E(0)}\big)\doteq E_{p}(T)$ is indeed the correction to the ground-state spectrum $E_{gs}(0)$ arising from weak coupling to the reservoir (see~\ref{sec:app1} for the proof of Eq.~\ref{eq:eq92}). Therefore, $E_{p}(T)$ could be taken into account as a perturbation, and could then be used to determine the temperature ranges over which the perturbative approach is valid. As a result, $E_{p}(T)\ll 1$. Because $E_{p}(T)$ is an absolutely system-dependent function of temperature with therefore no unique form, we consequently used plotting to determine the validity range for each of the systems of interest. Such a constraint could also be viewed as the self-consistency condition for Eq.~\ref{eq:eq9}, which imposes $\bra{\psi_{E}}\hat{H}_{I}\ket{\psi_{E}}$ must be equal to $\bra{\psi_{E}}\hat{H}_{I-1}\ket{\psi_{E}}$ within a certain numerical tolerance---$\ket{\psi_{E}}$ is the unperturbed eigenket of energy.
\section{\label{sec:4}Results and discussion}
\subsection{Free particle in a box}
\begin{figure}[H]
	\centering
	\fbox{\rule[0cm]{0cm}{0cm} \rule[0cm]{0cm}{0cm} 
	\subfigure[]{\label{subfig:2(a)}
		\includegraphics[scale=0.31,angle=-90]{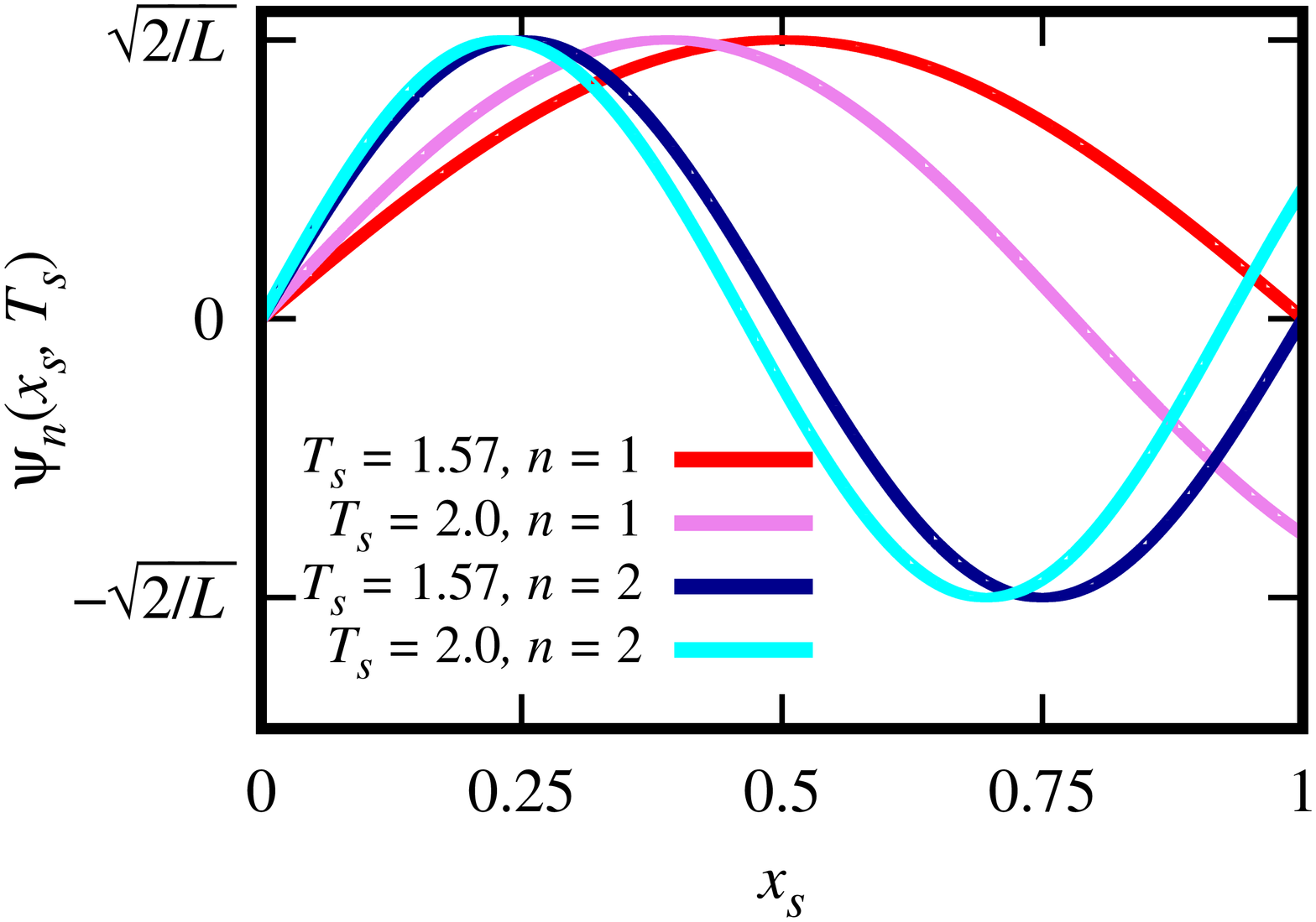}}
	\subfigure[]{\label{subfig:2(b)}
		\includegraphics[scale=0.31,angle=-90]{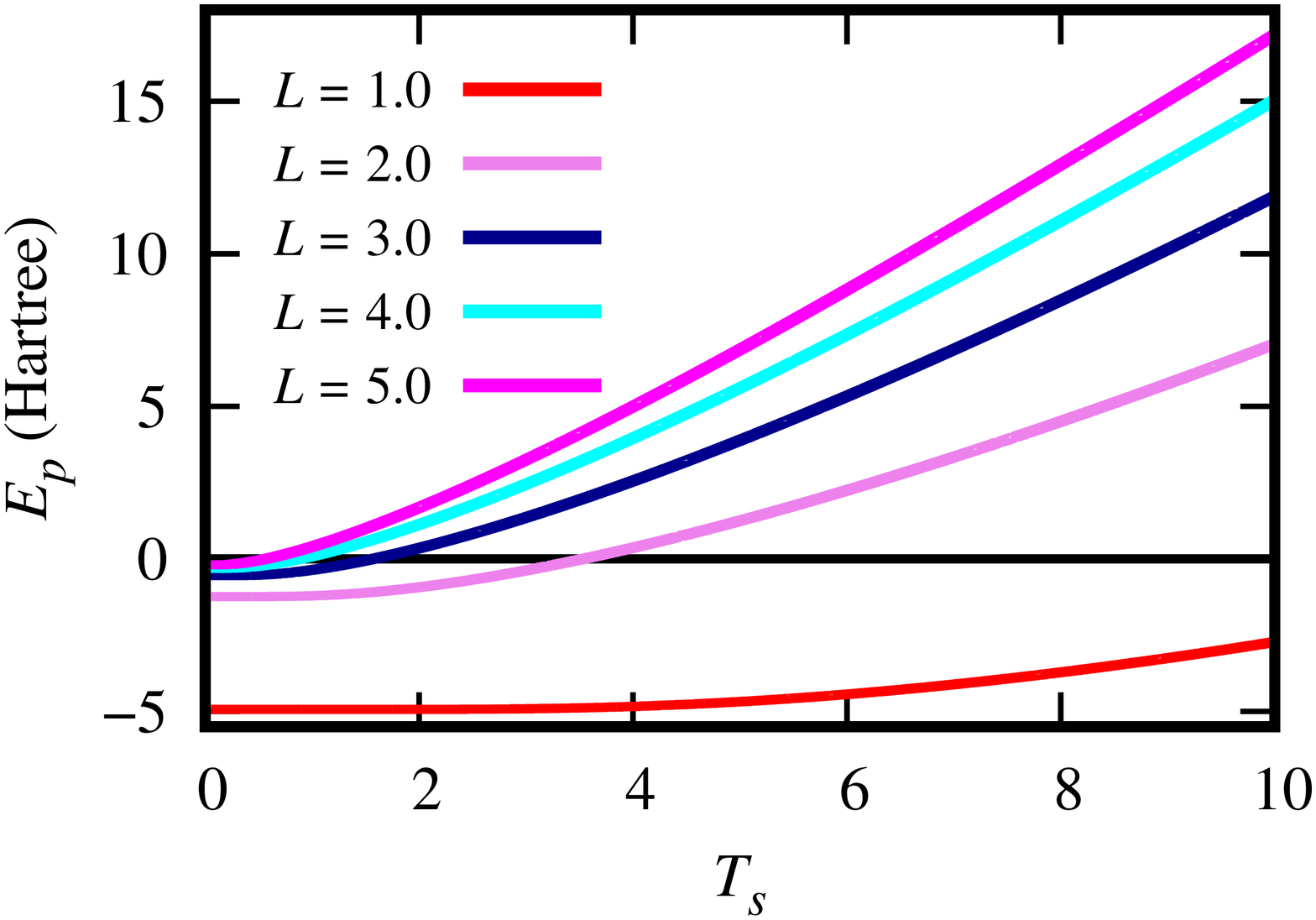}}}
	\caption{\label{fig:2}
		(a) The wavefunction of Eq.~\ref{eq:eq13} for the first two modes $n=1$ and 2 at $T_{s}=1.57$ and 2.0 within a box with $L=3.0$---rendered in Gnuplot (version 5.2)~\cite{46}. The $x_s$ ($=x\big/L$) and $T_s$ are respectively the scaled position and temperature in Hartree units. (b) The perturbation $E_{p}(T)=k_{B}T\ln\big(\sum_{n'=1}^{10}e^{-\beta\mathrm{\Xi} n'^{2}}\big)$ as a function of temperature for different values of $L$ from 1.0 to 5.0. This diagram exactly determines the ranges over which the perturbative approach is valid.}
\end{figure}

For a free, quantum particle with mass $m$ confined to move inside an infinite, one-dimensional potential well defined by $v(0\le x\le L)=0$ and $v(L<x<0)=\infty$, the weak coupling term [$E_{p}(T)$] up to the first order of self-consistency ($I=1$) corresponding with the most dominant term of $e^{-\beta\hat{H}_{I-1}}$ (in Eq.~\ref{eq:eq9}), when operates on energy eigenkets, results in
\begin{eqnarray}
\label{eq:eq10}
E_{p}(T)=\bra{\psi_{E}}k_{B}T\ln{\mathrm{Tr}}\left(e^{-\beta \hat{H}_{0}}\right)\ket{\psi_{E}}=k_{B}T\ln{\mathrm{Tr}}\Big(e^{-\beta E_{n'}(0)}\Big)=k_{B}T\ln\left(\sum_{n'=1}^{10}e^{-\beta\mathrm{\Xi} n'^{2}}\right),
\end{eqnarray} 
where $\mathrm{\Xi}\doteq\pi^{2}\hbar^{2}\big/2mL^{2}$, and $\mathrm{\Xi} n'^{2}=E_{n'}(0)$ is the ground-state (zero-temperature) energy of the system associated with state $n'$ (see~\ref{sec:app1} for the proof of Eq.~\ref{eq:eq10}). The summation over all states has also been approximated by the first ten states of the system. Putting Eq.~\ref{eq:eq10} back into Eq.~\ref{eq:eq92}, the energy spectrum of interest would then be $E_{n}(T)=E_{n}(0)+E_{p}(T)$, which, clearly reduces to the ground-state expression $E_{n}(0)=\mathrm{\Xi} n^{2}$ in the limit $T\longrightarrow 0$, as expected. The temperature-dependent wavefunction of the system would also be
\begin{eqnarray}
\label{eq:eq13}
\mathrm{\Psi}_{n}(x,T)=\sqrt{\frac{2}{L}}\sin\left(\frac{1}{\hbar}\sqrt{2mE_{n}(T)}\hspace{1mm}x\right),
\end{eqnarray}
which is illustrated in Fig.~\ref{subfig:2(a)} at two different temperatures including 1.57 and 2.0, and for the box size of 3.0, all in Hartree units. These temperature values are indeed within the validity range of the perturbative approach, as determined by Fig.~\ref{subfig:2(b)}, showing $E_{p}(T)$ for different values of the box size from $L=1.0$ to 5.0.

Evidently, Eq.~\ref{eq:eq13} also reduces to its corresponding zero-temperature analogue, $\mathrm{\Psi}_{n}(x)=\sqrt{2/L}\sin\big(n\pi x\big/L\big)$, when $T\longrightarrow 0$. However, there exists a serious discrepancy between $\mathrm{\Psi}_{n}(x,T)$ and $\mathrm{\Psi}_{n}(x)$. As seen in Fig.~\ref{subfig:2(a)}, $T_{s}=1.57$ is the temperature at which the perturbation is zero [according to Fig.~\ref{subfig:2(b)}], and therefore, the associated wavefunctions coincide with their corresponding zero-temperature counterparts as they vanish at the boundaries ($x=0$ and $L$). For $T_{s}=2.0$, on the other hand, the two wavefunctions ($n=1,2$) do not vanish at $x=L$ (i.e., $x_{s}=1$), which is clearly a chief consequence of weak coupling to the reservoir---such an effect would have indeed been observed at both boundaries if, in the numerical calculations, the box was defined symmetrically from $-L\big/2$ to $+L\big/2$, for instance. 

\begin{figure}[H]
	\centering
	\fbox{\rule[0cm]{0cm}{0cm} \rule[0cm]{0cm}{0cm} 
	\includegraphics[scale=0.32,angle=-90]{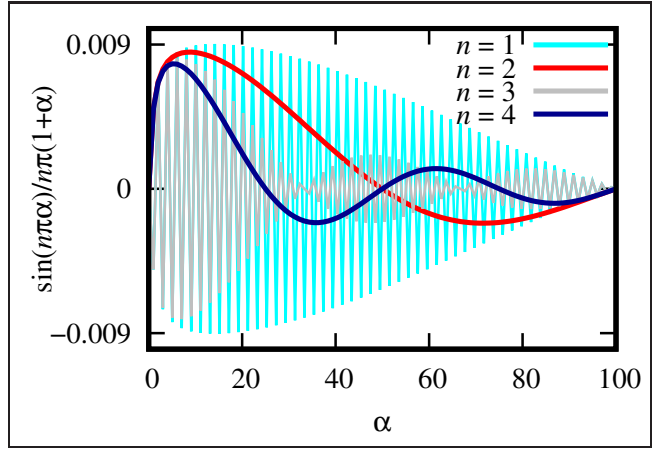}}
	\caption{\label{fig:3}
	The last term of Eq.~\ref{eq:eq133}, $\sin(n\pi\alpha)\big/n\pi(1+\alpha)$, as a function of $\alpha$ for the first four modes ($n=1,...,4$). For our purposes, $0<\alpha\ll 1$, however, this so-called residual probability takes very small values within zero to $\pm 0.009$ over the entire interval of $0\le\alpha\le100$.}
\end{figure}

One of the major consequences of this $\mathrm{\Psi}_{n}(x=0{\hspace{2mm}}{\mathrm{or}}{\hspace{2mm}}L,T)$ is that the probability of finding the particle somewhere within the system is not exactly equal to unity, or more precisely:
\begin{equation}
\label{eq:eq133}
\int_{-L/2}^{L/2}dx\Big|\mathrm{\Psi}_{n}(x,T)\Big|^{2}=1-\frac{\sin\big(n\pi \alpha\big)}{n\pi\big(1+\alpha\big)},
\end{equation}
where $\alpha\doteq E_{p}(T)\big/2E_{n}(0)$ [see \ref{sec:app3} for the proof of Eq.~\ref{eq:eq133}]. The last term is clearly the effect of weak coupling to the reservoir and its variation with respect to $\alpha$ is shown in Fig.~\ref{fig:3}.
As is seen, because $\alpha\ll 1$ (due to perturbation), $\lim_{\alpha\rightarrow 0}\frac{\sin(n\pi \alpha)}{n\pi(1+\alpha)}\longrightarrow0$, showing that the probability of finding the particle within the system is very close to unity. Even, for $\alpha>1$, which falls outside the present framework, this residual probability oscillates between zero and $\pm 0.009$, still being far less than 1. Nevertheless, this term inevitably means as the probability of finding the particle out of the system, or equivalently, within the reservoir (surrounding), which, in turn, implies some kind of quantum-mechanical tunneling as a consequence of weak coupling to the environment. This is indeed an expected effect, because, once the walls are impenetrable, no interaction between the system and the reservoir, even in the form of weak coupling, could then take place. Nonetheless, Eq.~\ref{eq:eq13} conforms to all the necessary criteria for quantum-mechanical wavefunctions including being finite everywhere, as well as being square-integrable. The orthogonality of $\mathrm{\Psi}_{n}(x,T)$ has also been proved in~\ref{sec:app2}. 
\subsection{Free particle in vacuum}
For a free, quantum particle with a well-defined linear momentum $p=\hbar k$ ($k$ being the wave number), we have
\begin{eqnarray}
\label{eq:eq14}
E_{p}(T)=\bra{p}k_{B}T\ln{\mathrm{Tr}}\left(e^{-\beta \hat{H}_{0}}\right)\ket{p}=\bra{p}\int_{-\infty}^{\infty}dp\ket{p}\bra{p}k_{B}T\ln{\mathrm{Tr}}\left(e^{-\beta \hat{H}_{0}}\right)\ket{p}\nonumber\\=k_{B}T\ln\int_{-\infty}^{\infty}dp\hspace{1mm}e^{-\beta p^{2}\big/2m}=k_{B}T\ln\sqrt{2\pi mk_{B}T},
\end{eqnarray}
where we have used the completeness condition for the simultaneous eigenket $\ket{p}$ of energy and momentum. Inserting Eq.~\ref{eq:eq14} in Eq.~\ref{eq:eq92}, it is obtained that
\begin{equation*}
\label{eq:eq15-1}
E_{k}(T)=\frac{\hbar^{2}k^{2}}{2m}+\langle E\rangle\ln\big(4\pi m\langle E\rangle\big),
\end{equation*}
where $\langle E\rangle=k_{B}T\big/2$ from the equipartition theorem, and $\lim_{T\rightarrow 0}E_{k}(T)=(\hbar^{2}\big/2m)k^{2}=E_{k}(0)$, which is the ground-state spectrum. The wavefunction of the system is therefore simply
\begin{equation}
\label{eq:eq16}
\mathrm{\Psi}_{k}^{\pm}(x,T)=e^{\pm ik(T)x},
\end{equation}
where
\begin{equation}
\label{eq:eq161}
k(T)=\bigg[k^{2}+\left(\frac{m}{\hbar^{2}}\right)k_{B}T\ln\big(2\pi mk_{B}T\big)\bigg]^{1/2}
\end{equation}
describes the square modulus of the temperature-dependent wave vector, and clearly reduces to $k^{2}$ in the limit $T\longrightarrow 0$. The $+\big/-$ signs in Eq.~\ref{eq:eq16} also denote traveling to right$\big/$left. 

\begin{figure}[H]
	\centering
	\fbox{\rule[0cm]{0cm}{0cm} \rule[0cm]{0cm}{0cm} 
	\includegraphics[scale=0.32,angle=-90]{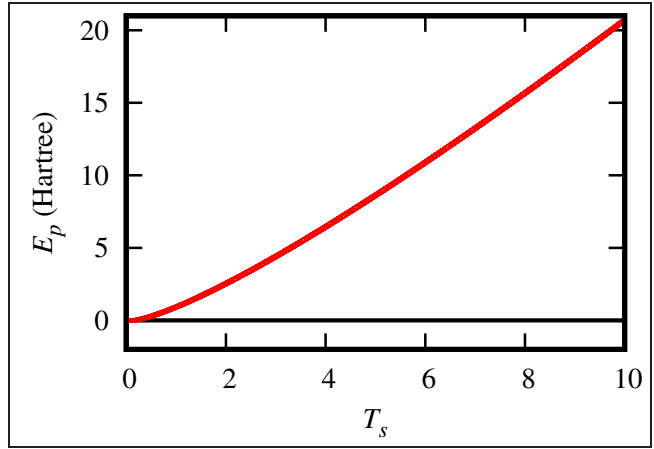}}
	\caption{\label{fig:4}
		The perturbation $E_{p}(T)=\langle E\rangle\ln\big(4\pi m\langle E\rangle\big)$ as a function of temperature, which is vanishing only at absolute zero. The $T_s$ is the scaled temperature in Hartree units.}
\end{figure}

\begin{figure}[H]
	\centering
	\fbox{\rule[0cm]{0cm}{0cm} \rule[0cm]{0cm}{0cm} 
	\subfigure[]{\label{subfig:5(a)}
		\includegraphics[scale=0.31,angle=-90]{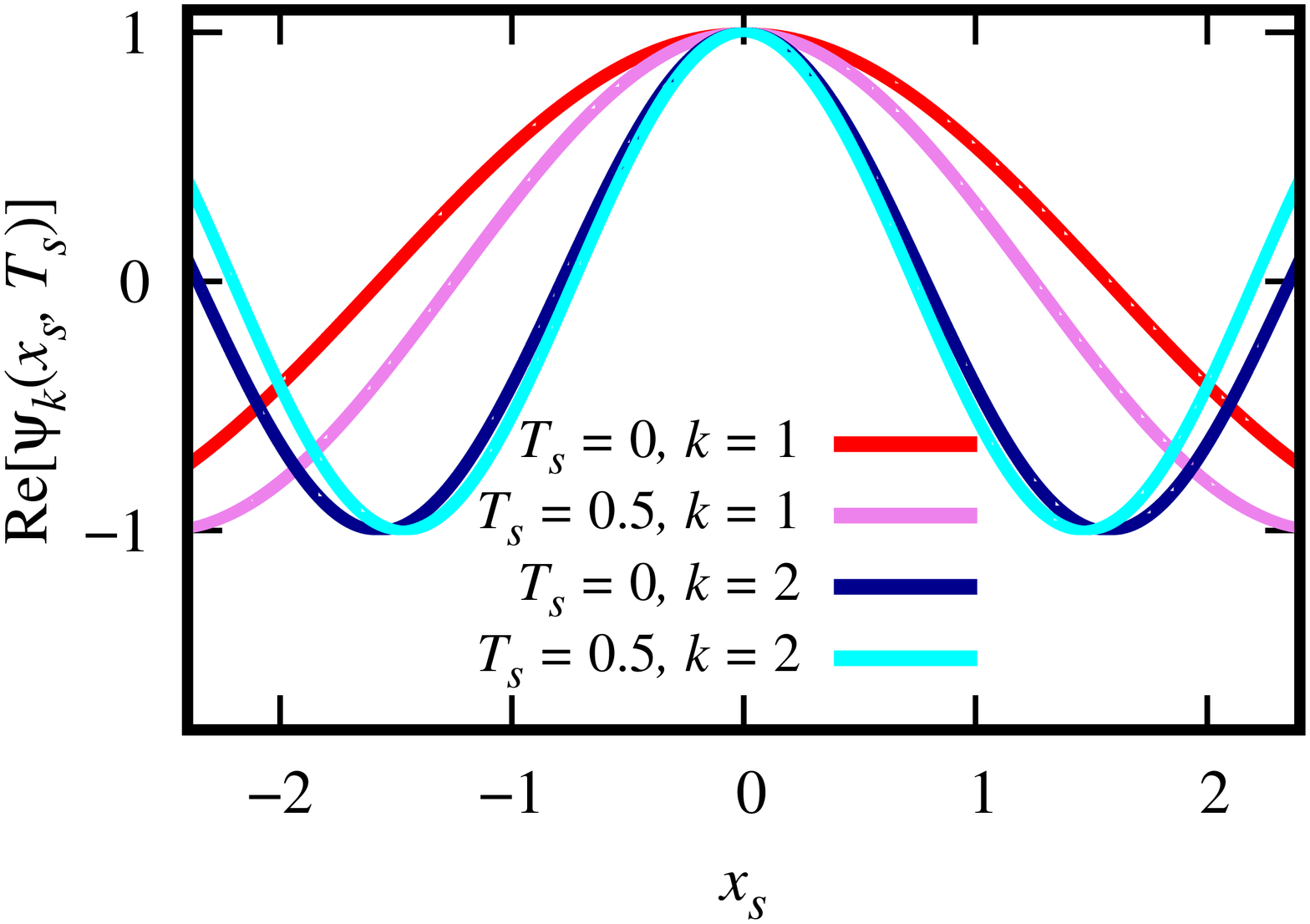}}
	\subfigure[]{\label{subfig:5(b)}
		\includegraphics[scale=0.31,angle=-90]{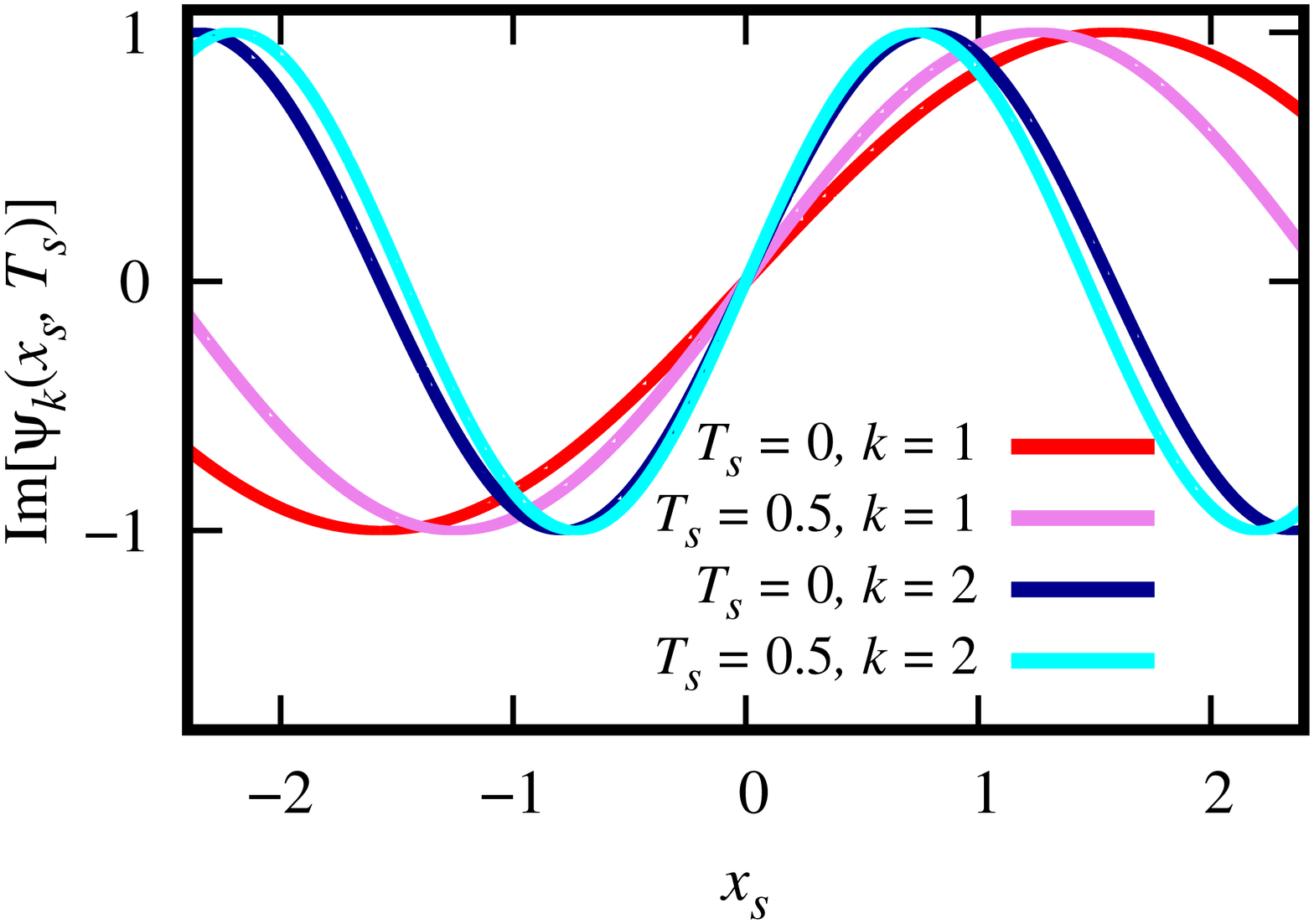}}}
	\caption{\label{fig:5}
		(a) The real (Re), and (b) the imaginary (Im) parts of the wavefunction Eq.~\ref{eq:eq16} for a free particle moving to right with wave numbers $k=1$ and 2 at $T_{s}=0$ and 0.5. The $x_s$ and $T_s$ are the scaled position and temperature, respectively.}
\end{figure}

The $E_{p}(T)$ vanishes only at absolute zero according to Fig.~\ref{fig:4}, and therefore, temperature values close enough to zero are only allowed.
Fig.~\ref{fig:5} shows the real (Re) and imaginary (Im) parts of Eq.~\ref{eq:eq16} for $k=1$ and 2 at two different temperatures including absolute zero.
In the same way as the particle in a box, the effect of temperature manifests in the form of some contraction in the real and imaginary parts of the corresponding zero-temperature wavefunction of the system. This could indeed be inferred by rewriting Eq.~\ref{eq:eq161} using the binomial series. As a result,
\begin{equation*}
\label{eq:eq162}
k(T)\simeq k\bigg[1+\left(\frac{mk_{B}T}{2\hbar^{2}k^{2}}\right)\ln\big(2\pi mk_{B}T\big)\bigg],
\end{equation*}
from which Eq.~\ref{eq:eq16} turns into 
\begin{equation*}
\label{eq:eq163}
\mathrm{\Psi}_{k}^{\pm}(x,T)\simeq e^{\pm ikx}e^{\pm i\Big[\frac{mk_{B}T}{2\hbar^{2}k}\ln(2\pi mk_{B}T)\Big]x},
\end{equation*}
where the second exponential is clearly so close to unity, showing that the behavior of Eq.~\ref{eq:eq16} must necessarily be very similar to the ground-state counterpart ($e^{\pm ikx}$).
\subsection{Harmonic oscillator}
\begin{figure}[H]
	\centering
	\fbox{\rule[0cm]{0cm}{0cm} \rule[0cm]{0cm}{0cm} 
	\includegraphics[scale=0.32,angle=-90]{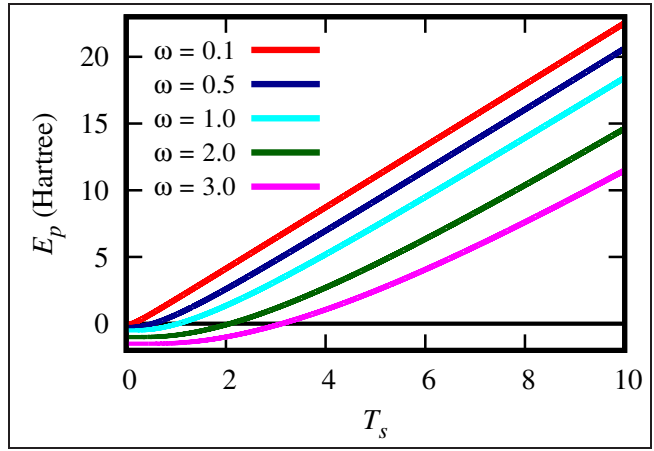}}
	\caption{\label{fig:6}
		The weak-coupling term $E_{p}(T)=k_{B}T\ln\left(\sum_{n'=0}^{9}e^{-\beta(n'+1/2)\hbar\omega}\right)$ as a function of temperature for the five typical values of the angular frequency ($\omega$). The zero of this term, showing the validity range of the perturbative approach, takes place at higher temperature values as $\omega$ increases. All the variables are in Hartree units.}
\end{figure}

\begin{figure}[H]
	\centering
	\fbox{\rule[0cm]{0cm}{0cm} \rule[0cm]{0cm}{0cm} 
	\includegraphics[scale=0.625,angle=-90]{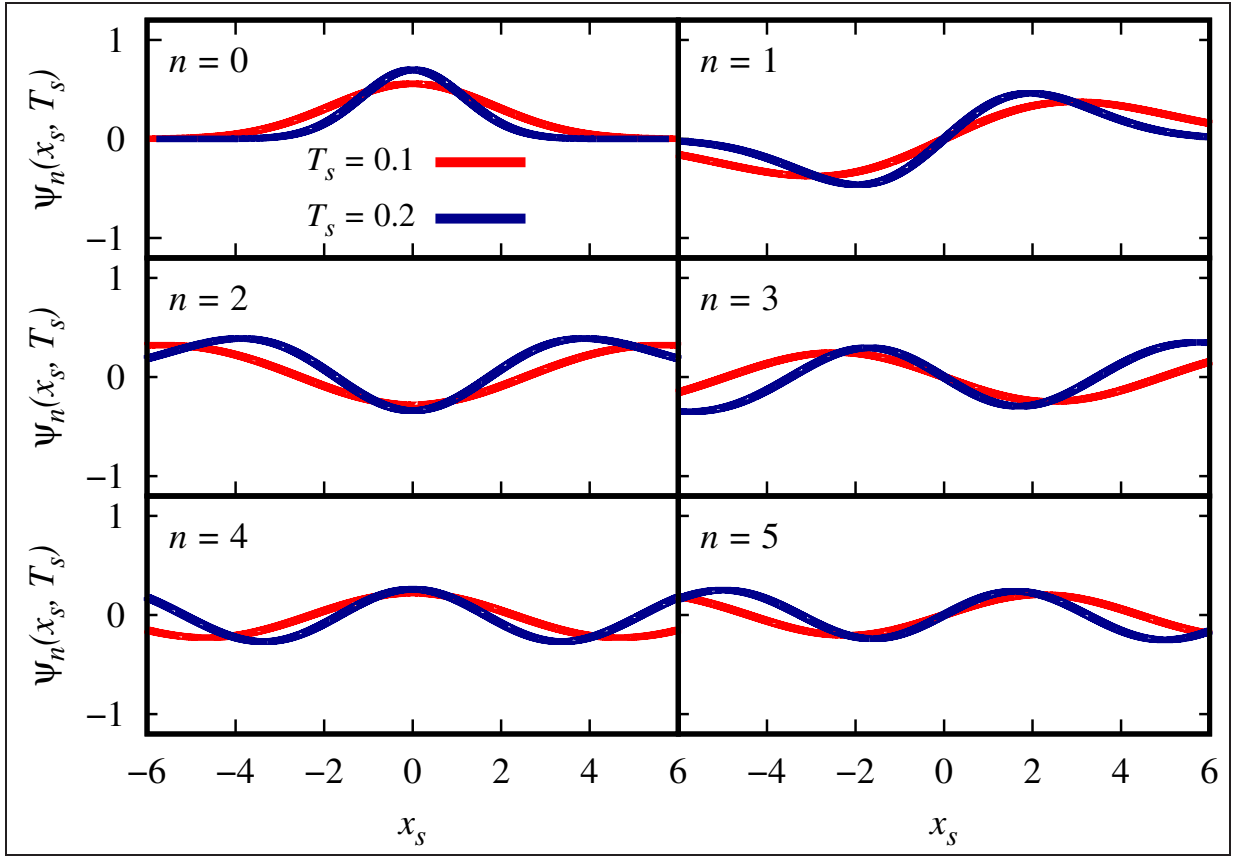}}
	\caption{\label{fig:7}
		The first six modes, $n=0$ to 5, of the wavefunction Eq.~\ref{eq:eq21} at $T_{s}=0.1$ and 0.2. Like the two preceding systems, the effect of weak coupling emerges in the form of slight contractions in the associated ground-state wavefunctions. The $x_{s}$ $(=x\big/x_{0})$ and $T_s$ are the scaled position and temperature, respectively.}
\end{figure}

For a single quantum harmonic oscillator with mass $m$, angular frequency $\omega$, and one-dimensional harmonic potential $m\omega^{2}x^{2}\big/2$, we then have
\begin{eqnarray}
\label{eq:eq17}
E_{p}(T)=\bra{n'}k_{B}T\ln{\mathrm{Tr}}\left(e^{-\beta \hat{H}_{0}}\right)\ket{n'}=k_{B}T\ln{\mathrm{Tr}}\left(e^{-\beta\hbar\omega(n'+1/2)}\right)=k_{B}T\ln\left(\sum_{n'=0}^{9}e^{-\beta\hbar\omega(n'+1/2)}\right),
\end{eqnarray} 
where $\big(n'+1\big/2\big)\hbar\omega=E_{n'}(0)$ is the corresponding ground-state energy of the $n'$th state, and the sum-over-all-states is approximated by the first ten states. $\ket{n'}$ also denotes the unperturbed eigenket of energy. Inserting Eq.~\ref{eq:eq17} back into Eq.~\ref{eq:eq92}, then
\begin{eqnarray*}
\label{eq:eq19}
E_{n}(T)\doteq\bigg(n+\frac{1}{2}\bigg)\hbar\mathrm{\Omega}_{n}(T)= \bigg(n+\frac{1}{2}\bigg)\hbar\omega+k_{B}T\ln\left(\sum_{n'=0}^{9}e^{-\beta(n'+1/2)\hbar\omega}\right),
\end{eqnarray*}
which is the associated temperature-dependent energy spectrum of the oscillator. The generalized, temperature-dependent angular frequency would then be
\begin{equation}
\label{eq:eq20}
\mathrm{\Omega}_{n}(T)=\omega+\frac{k_{B}T}{\big(n+1/2\big)\hbar}\ln\left(\sum_{n'=0}^{9}e^{-\beta(n'+1/2)\hbar\omega}\right),
\end{equation}
Evidently, Eq.~\ref{eq:eq20} results in the ground-state spectrum $E_{n}(0)=\big(n+1\big/2\big)\hbar\omega$ in the limit $T\longrightarrow 0$. The associated wavefunction is consequently
\begin{equation}
\label{eq:eq21}
\mathrm{\Psi}_{n}(x,T)=\bigg(\sqrt{\pi}\hspace{1mm}2^{n}\hspace{1mm}n!\hspace{1mm}x_{0}(n,T)\bigg)^{-1/2}e^{-x^{2}\big/2x_{0}^{2}(n,T)}\hspace{1mm}\mathcal{H}_{n}\left[\frac{x}{x_{0}(n,T)}\right],
\end{equation}
where $x_{0}(n,T)=\big(\hbar\big/m\mathrm{\Omega}_{n}\big)^{1/2}$ defines the characteristic, temperature-dependent length scale of the oscillator, ans $\mathcal{H}_{n}$ is the Hermite polynomial of degree $n$. The validity range for the equilibrium temperature could be determined by the zeros of $E_{p}(T)$, as seen in Fig.~\ref{fig:6}.
Fig.~\ref{fig:7} also illustrates the first six modes ($n=0,...,5$) of the wavefunction Eq.~\ref{eq:eq21} at two different temperatures, which are allowed according to Fig.~\ref{fig:6}.
As is seen, in the same way as the two preceding systems, the effect of temperature (weak coupling) is in the form of some contractions in the corresponding zero-temperature wavefunctions of the oscillator in a way that increase in temperature directly results in higher degree of compactness. Such an effect accordingly makes the wavefunctions to be non-vanishing at the two boundaries, as a chief consequence of weak coupling to thermal reservoir. 

However, the wavefunctions of Eq.~\ref{eq:eq21} are orthogonal, also the probability of finding the particle within the system is exactly equal to unity (see~\ref{sec:app4} for the proof of orthonormality), in contrast with the case of a free particle in a box, in which the particle could be found, though by a very small probability, within the reservoir. This was indeed expected based on the intuition that tunneling could only take place for free particles, and mathematically due to the fact that the integration interval for the probability of interest covers the entire position axis from $-\infty$ to $\infty$.

The two criteria of being finite everywhere, as well as being square-integrable are also be conformed, as inferred from Fig.~\ref{fig:7}.
\section{\label{sec:5}Conclusions}
We adopted the perturbative approach in order to advance a temperature-dependent version of quantum mechanics in the non-relativistic regime at low-enough temperatures. Using the laws of thermodynamics, we accordingly derived a generalized, self-consistent Hamiltonian for a given quantum system, involving a temperature-dependent term, which connects the system of interest and its immediate environment, and reduces to its ground-state counterpart at absolute zero. The weak-coupling term was therefore considered as the perturbation exerted to the system. Applying our formalism to some quantum-mechanical systems having exact zero-temperature solutions, including the free particle in a box, the free particle in vacuum, and the harmonic oscillator, up to the first order of self-consistency, accordingly led to the temperature-dependent Hamiltonian, energy spectrum, and wavefunction for each of the systems, consistent with the limit of low temperatures. As a chief consequence of coupling to the thermal reservoir, some kind of quantum-mechanical tunneling effect by a residual probability was also inferred for the free particle in a box. We further investigated and discussed the effects of weak coupling on the main properties of the wavefunctions for the aforementioned systems in terms of orthonormality, being finite everywhere, being square-integrable, and being vanishing at the boundaries.
\appendix
\section{Proof of Eq.~\ref{eq:eq92}}
\label{sec:app1}
From Eq.~\ref{eq:eq9} and for $I=1$, we then have
\begin{eqnarray}
\label{eq:eqap}
\bra{\psi_{E'}}\hat{H_{1}}\ket{\psi_{E'}}=\bra{\psi_{E'}}\hat{H}_{0}\ket{\psi_{E'}}+\bra{\psi_{E'}}k_{B}T\ln{\mathrm{Tr}}\left(e^{-\beta\hat{H}_{0}}\right)\ket{\psi_{E'}}
\Longrightarrow E(T)=E(0)+E_{p}(T),
\end{eqnarray}
where
\begin{eqnarray}
E_{p}(T)=\bra{\psi_{E'}}k_{B}T\ln{\mathrm{Tr}}\left(e^{-\beta\hat{H}_{I-1}}\right)_{I=1}\ket{\psi_{E'}}=
\bra{\psi_{E'}}k_{B}T\ln{\mathrm{Tr}}\left(e^{-\beta \hat{H}_{0}}\right)\ket{\psi_{E'}}\nonumber\\=\bra{\psi_{E'}}k_{B}T\ln{\mathrm{Tr}}\left(e^{-\beta \hat{H}_{0}}\hat{\mathcal{I}}\right)\ket{\psi_{E'}}
=\bra{\psi_{E'}}k_{B}T\ln{\mathrm{Tr}}\left(e^{-\beta \hat{H}_{0}}\sum_{E'}\ket{\psi_{E'}}\bra{\psi_{E'}}\right)\ket{\psi_{E'}}\nonumber\\
=\bra{\psi_{E'}}k_{B}T\ln{\mathrm{Tr}}\left(\sum_{E'}e^{-\beta \hat{H}_{0}}\ket{\psi_{E'}}\bra{\psi_{E'}}\right)\ket{\psi_{E'}}
=\bra{\psi_{E'}}k_{B}T\ln{\mathrm{Tr}}\left(\sum_{E'}e^{-\beta E_{n'}(0)}\ket{\psi_{E'}}\bra{\psi_{E'}}\right)\ket{\psi_{E'}}\nonumber\\
=\bra{\psi_{E'}}k_{B}T\ln{\mathrm{Tr}}\left(e^{-\beta E_{n'}(0)}\sum_{E}\ket{\psi_{E'}}\bra{\psi_{E'}}\right)\ket{\psi_{E'}}
=k_{B}T\ln{\mathrm{Tr}}\bigg(e^{-\beta E_{n'}(0)}\bigg)\braket{\psi_{E'}|\psi_{E'}}\nonumber\\=k_{B}T\ln{\mathrm{Tr}}\bigg(e^{-\beta E_{n'}(0)}\bigg)\nonumber,
\end{eqnarray}
in which we have used the completeness condition $\sum_{E'}\ket{\psi_{E'}}\bra{\psi_{E'}}=\hat{\mathcal{I}}$ for energy eigenkets. The perturbed wavefunction of the system in Eq.~\ref{eq:eqap} has also been approximated by the unperturbed one (i.e., up to the zeroth order of perturbation).\\
\section{Orthogonality of Eq.~\ref{eq:eq13}}
\label{sec:app2}
We have
\begin{eqnarray}
\label{eq:app2}
E_{n}(T)=E_{n}(0)+E_{p}(T)=\frac{n^{2}\pi^{2}\hbar^{2}}{2mL^{2}}+E_{p}(T)
\Longrightarrow \sqrt{\frac{2mE_{n}(T)}{\hbar^{2}}}=\frac{n\pi}{L}\sqrt{1+\frac{E_{p}(T)}{E_{n}(0)}}\nonumber\\\simeq\frac{n\pi}{L}\left(1+\frac{1}{2}\frac{E_{p}(T)}{E_{n}(0)}\right)
\doteq\frac{n\pi}{L}\big(1+\alpha\big),
\end{eqnarray}
where we have used the binomial series. Plugging Eq.~\ref{eq:app2} back into Eq.~\ref{eq:eq13} accordingly results in
\begin{eqnarray}
\label{eq:app3}
\int_{-L/2}^{L/2}dx\hspace{1mm}\mathrm{\Psi}_{m}(x,T)\mathrm{\Psi}_{n}(x,T)=\frac{2}{L}\int_{-L/2}^{L/2}dx\hspace{1mm}\sin\bigg(\frac{m\pi}{L}x\big(1+\alpha\big)\bigg)\sin\bigg(\frac{n\pi}{L}x\big(1+\alpha\big)\bigg)\nonumber\\
=\frac{2}{L\big(1+\alpha\big)}\int_{-L\big(1+\alpha\big)\big/2}^{L\big(1+\alpha\big)\big/2}dX\hspace{1mm}\sin\bigg(\frac{m\pi}{L}X\bigg)\sin\bigg(\frac{n\pi}{L}X\bigg)
=\frac{2}{L\big(1+\alpha\big)}\frac{L}{2}\big(1+\alpha\big)\delta_{mn}=\delta_{mn},\nonumber
\end{eqnarray}
where we have applied the change of variable $x\big(1+\alpha\big)=X$, and $\delta_{mn}$ is the Kronecker delta.
\section{Proof of Eq.~\ref{eq:eq133}}
\label{sec:app3}
Using Eq.~\ref{eq:eq13}, the probability of interest would then be
\begin{eqnarray}
\int_{-L\big/2}^{L\big/2}dx\hspace{1mm}\Big|\mathrm{\Psi}_{n}(x,T)\Big|^{2}=\frac{2}{L}\int_{-L\big/2}^{L\big/2}dx\hspace{1mm}\sin^{2}\bigg(\frac{n\pi}{L}x\big(1+\alpha\big)\bigg)
=\frac{1}{L}\int_{-L\big/2}^{L\big/2}dx\hspace{1mm}\Bigg[1-\cos\left(\frac{2n\pi}{L}x\big(1+\alpha\big)\right)\Bigg]\nonumber\\=1-\frac{\sin\left[n\pi\big(1+\alpha\big)\right]}{n\pi\big(1+\alpha\big)}
=1-\frac{\sin\big(n\pi\alpha\big)}{n\pi\big(1+\alpha\big)}.\nonumber
\end{eqnarray}
\section{Orthonormality of Eq.~\ref{eq:eq21}}
\label{sec:app4}
We can write
\begin{eqnarray}
\label{eq:app5}
x_{0}(n,T)=\bigg(\frac{\hbar}{m\Omega_{n}}\bigg)^{1/2}=\Bigg(\frac{\hbar}{m\omega+\frac{mE_{p}(T)}{(n+1/2)\hbar}}\Bigg)^{1/2}=x_{0}\Bigg(1+\frac{E_{p}(T)}{E_{n}(0)}\Bigg)^{-1/2}\simeq x_{0}\Bigg(1-\frac{E_{p}(T)}{2E_{n}(0)}\Bigg)\nonumber\\=x_{0}\big(1-\alpha\big),
\end{eqnarray}
where we have used the binomial series, and $x_{0}=(\hbar\big/m\omega)^{1/2}$ is the ground-state length scale of the oscillator. Therefore, using Eqs.~\ref{eq:eq21} and \ref{eq:app5}, we would have
\begin{eqnarray}
\int_{-\infty}^{\infty}dx\hspace{1mm}\mathrm{\Psi}_{m}^{*}(x,T)\mathrm{\Psi}_{n}(x,T)=\int_{-\infty}^{\infty}dx\hspace{1mm}\bigg(\sqrt{\pi}\hspace{1mm}2^{m}m!\hspace{1mm}x_{0}\bigg)^{-1/2}\bigg(\sqrt{\pi}\hspace{1mm}2^{n}n!\hspace{1mm}x_{0}\bigg)^{-1/2}\nonumber\\\times\bigg(1+\frac{\alpha}{2}\bigg)^{2}e^{-x^2\big/[x_{0}(1-\alpha)]^2}\mathcal{H}_{m}\left[\frac{x}{x_{0}\big(1-\alpha\big)}\right]\mathcal{H}_{n}\left[\frac{x}{x_{0}\big(1-\alpha\big)}\right]\nonumber\\
=\bigg(\sqrt{\pi}\hspace{1mm}2^{m}m!\hspace{1mm}x_{0}\bigg)^{-1/2}\bigg(\sqrt{\pi}\hspace{1mm}2^{n}n!\hspace{1mm}x_{0}\bigg)^{-1/2}\bigg(1+\frac{\alpha}{2}\bigg)^{2}\nonumber\\
\times\int_{-\infty}^{\infty}dx\hspace{1mm}\mathcal{H}_{m}\left[\frac{x}{x_{0}\big(1-\alpha\big)}\right]\mathcal{H}_{n}\left[\frac{x}{x_{0}\big(1-\alpha\big)}\right]e^{-x^2\big/[x_{0}(1-\alpha)]^2}\nonumber\\
=\bigg(\sqrt{\pi}\hspace{1mm}2^{m}m!\hspace{1mm}x_{0}\bigg)^{-1/2}\bigg(\sqrt{\pi}\hspace{1mm}2^{n}n!\hspace{1mm}x_{0}\bigg)^{-1/2}\bigg(1+\frac{\alpha}{2}\bigg)^{2}x_{0}\big(1-\alpha\big)\nonumber\\
\times\int_{-\infty}^{\infty}d\left[\frac{x}{x_{0}\big(1-\alpha\big)}\right]\mathcal{H}_{m}\left[\frac{x}{x_{0}\big(1-\alpha\big)}\right]\mathcal{H}_{n}\left[\frac{x}{x_{0}\big(1-\alpha\big)}\right]e^{-x^2\big/[x_{0}(1-\alpha)]^2}\nonumber\\
=\bigg(\frac{n!}{m!}\hspace{1mm}2^{n-m}\bigg)^{1/2}\bigg(1+\frac{\alpha}{2}\bigg)^{2}\big(1-\alpha\big)\hspace{1mm}\delta_{mn},\nonumber
\end{eqnarray}
which is the desired result. To further show the orthonormality of the wavefunctions, of course without loss of generality, we proceed by considering the case with $n=1$. As a result,
\begin{eqnarray}
\int_{-\infty}^{\infty}dx\hspace{1mm}\bigg|\mathrm{\Psi}_{1}(x,T)\bigg|^{2}=\int_{-\infty}^{\infty}dx\hspace{1mm}\Bigg|\bigg(\sqrt{\pi}\hspace{1mm}2x_{0}(1,T)\bigg)^{-1/2}e^{-x^2\big/2x_{0}^{2}(1,T)}\mathcal{H}_{1}\left[\frac{x}{x_{0}(1,T)}\right]\Bigg|^{2}\nonumber\\
=\int_{-\infty}^{\infty}dx\hspace{1mm}\Bigg|\bigg(\sqrt{\pi}\hspace{1mm}2x_{0}\big(1-\alpha\big)\bigg)^{-1/2}e^{-x^2\big/2x_{0}^{2}(1,T)}\mathcal{H}_{1}\left[\frac{x}{x_{0}\big(1-\alpha\big)}\right]\Bigg|^{2}\nonumber\\
=\frac{1}{(1-\alpha)^{3}}\int_{-\infty}^{\infty}dx\hspace{1mm}\Bigg|\bigg(\sqrt{\pi}\hspace{1mm}2x_{0}\bigg)^{-1/2}e^{-x^2\big/2x_{0}^{2}(1-\alpha)^2}\mathcal{H}_{1}\left(\frac{x}{x_{0}}\right)\Bigg|^{2}=1.\nonumber
\end{eqnarray}
\bibliographystyle{unsrt}

\end{document}